# On the Physical Cause and the Distance of Gamma Ray Bursts and Related Phenomena in the X-Rays and the Ultra-Violet

Ernst Karl Kunst
Im Spicher Garten 5
53639 Königswinter
Germany

**The modified Lorentz transformation of a distance-dependent special theory of relativity - which will be briefly summarized - predicts the possibility of superluminal velocity of very distantly moving material bodies to be connected with the generation of Cerencov radiation off the quantum vacuum. It is shown that vacuum Cerencov radiation due to the superluminal propagation of extraterrestrial spaceprobes in the interstellar space would account for all known properties of gamma ray bursts (GRBs) and the "afterglow" at lower frequencies. Distances and other parameter prove to be calculable and the theoretical results on these grounds to be in good accord with experiment.**

## Introduction

As widely known, were those bursts detected by the Vela military satellites in the late 1960s. Currently the most precise findings come from the Burst and Transient Source Experiment (BATSE) on board the Compton Gamma Ray Observatory (CGRO), which has collected nearly threethousand GRBs since 11 April 1991 at a rate of $\approx$ 0.8/day . The most important results, established by BATSE, are that the GRBs are distributed fairly uniformly on the sky, and their spatial distribution is finite. These results taken together rule out sources confined to the galactic disc and nearby sources, in any case if neutron stars or other known galactic objects are considered to be those sources. Therefore, it has been proposed GRBs to be of extra galactic origin. Far away galaxies and other cosmic objects as sources of the GRBs seem to explain their finite distribution, their obvious attenuation with increasing distance. The successful identifications made by the BeppoSAX satellite in X-rays, followed by the detection of counterparts in visual light as well as radio waves, seem to support this point of view, which further has been stressed by the measurement of absorption lines in the spectrum of some counterparts. Thus, GRBs seem to be put definitely into the cosmological distance.
On the other hand if GRBs are really at cosmological distances, this leads to the energy and to the compactness problem in the respective astrophysical models. Furthermore, there remain other puzzling open problems in GRB studies, as for instance the unseen "afterglows" in most cases or the obvious lack of "host galaxies". In the following the well-known properties of the GRBs are briefly summarized.

    1) Burst durations range from 10 ms to hundreds of seconds (s), with a distribution maximum at $\approx$ 30 s (see 5) and Fig. 1);

    2) GRBs occur randomly in time and position on the sky; They are

distributed isotropically in direction and do not repeat, which means that after one ceases, it vanishes entirely. In most cases no steady object remains detectable;

3) GRBs tend to have similiar non thermal energy spectra; Photon energies are between ≈ 10 keV and some GeV, typically peaking in the MeV region;

4) The distribution of burst brightness follows a -3/2 power law for a narrow range and falls off elsewhere which implies that the population is confined in space with the distance limit unknown [1];

5) A subclass of short GRBs seems to be distributed anisotropically [2];

6) Burst durations show evidence for a bimodal distribution with a first event maximum at ≈ 0.3 - 0.4 s, a minimum at 2 s and an absolute maximum of burst events at ≈ 30 s [3]. But rather it seems to be a more trimodal distribution if attention is given to the but only apparently small second maximum at ≈ 1.35 s. If in Fig. 1 onto the latter the same scale is applied as valid for the maximum at ≈ 0.4 s, it becomes evident that it spans over a duration time of ≈ 0.5 s as compared with only ≈ 0.25 s of the first maximum;

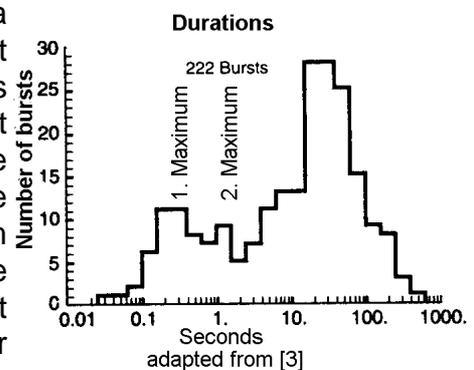

Fig. 1

7) Some light curves of GRBs show significant count rate variations on time scales as short as ms;

8) It also has been found that there may exist GRBs the soft emission of which has a time delay relative to the high energy emission [4];

9) The range of intensities for all GRBs ever detected covers a factor of ≈ 10000, from $10^{-3}$ erg/cm$^2$ down to $10^{-7}$ erg/cm$^2$;

10) GRBs seem to rise rapidly to a peak of intensity and then decay exponentially, which phenomenon is known as "FRED" (Fast Rise, Exponential Decay) [3];

11) In the present number of GRBs with X-ray prompt counterpart detected by BeppoSAX none is classified as short (sub second) event [5];

12) In the light of the afterglow of some GRBs a system of absorption lines has been found - first, with z = 0.835 in the spectrum of the optical counterpart to GRB 970508 - which show remarkably small velocity dispersion, suggesting that the lines arise from a single cloud [6].

## Superluminal Propagation of Material Bodies in Vacuo as the Cause of Cerencov Radiation

In the previous study [7] on relativistic kinematics among others has been shown that the transformations

$$x_2' = \gamma_0^* \left( x_1^* - \frac{v_0 t_1^*}{R} \right), \quad y_2' = y_1^*, \quad z_2' = z_1^*, \quad t_2' = \gamma_0^* \left( t_1^* - \frac{v_0 x_1^*}{c^2 R} \right),$$

$$x_1^{\circ *} = \gamma_0^* \left( x_2' + \frac{v_0 t_2'}{R} \right), \quad y_1^{\circ *} = y_2', \quad z_1^{\circ *} = z_2', \quad t_1^{\circ *} = \gamma_0^* \left( t_2' + \frac{v_0 x_2'}{c^2 R} \right),$$

are valid between an inertial system $S_1^*$ considered to be at rest and a very distantly moving inertial system $S_2'$, where

$$\gamma_0^* = \frac{1}{\sqrt{1 - \frac{v_0^2}{c^2 R^2}}}, \quad \gamma_0 = \frac{1}{\sqrt{1 - \frac{v_0^2}{c^2}}}.$$

and

$$R = \overline{S_1^* S_2'}.$$

From this follows that if

$$|R| = |\overline{S_1^* S_2'}| \gg |c \Delta t|,$$

the velocity of an object resting in the coordinate source of $S_2'$ relative to $S_1^*$ must be

$$\frac{\Delta x_1^*}{\Delta t_1^*} = \frac{\Delta x_2'}{\Delta t_2'} \gamma_0$$

or, as expressed in the coordinates and the time parameter of the rest frame $S_1$ - the conventional system of special relativity of the observer considered at rest:

$$u_{x_1}^* = u_{x_1} \gamma_0 = v_0 \gamma_0 \, .$$

The dilation of time in $S_2'$ is compensated for by the symmetrical inertial (far range) velocity

$$V_0 = v_0 \gamma_0 \tag{1}$$

of this system relative to $S_1^*$ also implying the velocity of light in $S_2'$ to be $C = c\gamma_0$ - if both systems are spatial far apart. As a result if $S_2'$ approaches $S_1^*$ observers resting in the coordinate sources of either system will meet or receive light signals emitted from the respective other system after the same amount of time has elapsed:

$$\begin{aligned} V_0 \Delta_{t_1} &= v_0 \Delta_{t_2}, \\ C \Delta_{t_1} &= c \Delta_{t_2}. \end{aligned} \tag{1a}$$

Any velocity even exeeding that of light in any amount is possible, allowing in principle the superluminal propagation of solid bodies and thereby superluminal transfer of information between very distant systems. Furthermore, it has been found that the probability p to encounter virtual dipoles or photons for a particle traversing the fluctuating quantum vacuum at subluminal symmetrical velocity $V_0 < c$ according to the uncertainty relation is given by

$$p = \frac{\Delta E \Delta t}{h} = \frac{\Delta E}{h \Delta v} \geq \frac{V_0^4}{c^4} \leq 1 \, ,$$

Hence, as bodies propagate superluminally through vacuo they generate Cerencov radiation because the energy of virtual dipoles becomes real and stable as the far range velocity $V_0$ exceeds that of light and the above probability relation becomes

$$\begin{aligned} p &= \frac{h V_0^4}{E c^4} = 1 = \text{const}, \\ &= \frac{h \gamma_0^4}{E} = 1 = \text{const} \end{aligned} \tag{2}$$

if $V_0/c \gg 1$, where h means Planck's constant and $E = (2E_{phot})^{1/2}$ "natural energy" acording to symmetrically modified special relativity [8]; $E_{phot}$ denotes conventional photon energy. For this radiation the relation

$$\cos\alpha = \frac{c}{V_0} = \frac{1}{\gamma_0\beta_0} = \frac{1}{\sqrt[4]{v}},$$

has been proposed, where $V_0 \geq c$, $\beta_0 = v_0/c$, $\alpha$ halfangle of the radiation cone and $v$ frequency of the radiated photon.
But subsequently I come to the conclusion that this derivation of the vacuum Cerenkov relation is incomplete and to begin with must be corrected to

$$2\sqrt{2}\cos\alpha = \frac{c}{V_0} = \frac{1}{\gamma_0\beta_0} = \frac{1}{\sqrt[4]{v}},$$
$$= \frac{1}{\gamma_0} \quad (3)$$

if $\beta_0 \to 1$. The reason is that $\sin\alpha$ of the Cerencov angle of a very distant superluminally moving system $S_2'$ must equal the relation $c/V_0 = c/(v_0\gamma_0)$. According to (2) is the generation of Cerenkov radiation of lowest energy to expect if $c/V_0 = c/(v_0\gamma_0) = 1$ implying $v_0/c = 1/\sqrt{2}$ and, thus, $\sin\alpha \leq 1$. Furthermore, according to (1a) the radiation of Cerenkov light of lowest energy will be observed by an observer based at $S_2'$ if $v_0/c \geq 1/\sqrt{2}$ and of highest energy if $v_0/c \to 1$. This implies that the radiation can only occur within a cone of $\alpha = \pi/4$. Accordingly an observer based at $S_1^*$ will this radiation observe in a cone ranging from $\alpha \geq 0$ to $\alpha \to \pi/4$.
But (3) can be valid only if $2\sqrt{2}\cos\alpha \to 1$ and $V_0$ only slightly $> c$ for the following reasoning.
Suppose the very distant system $S_2'$ is heading toward the observer resting in the coordinate source of $S_1^*$ at rest. As observed from the latter will all dimensions normal to the velocity vector of $S_2'$ remain unaltered. We introduce a system $S_1$ at rest relative to $S_1^*$ and also very far from $S_2'$ implying the latter to move parallel relative to the first. As observed from $S_1$ must (1a) be valid and, thus, the space-time of $S_2'$ elongated by a factor of $\gamma_0$. Accordingly the angle under which the Cerenkov light from the path of the superluminally moving system $S_2'$ is emitted toward the resting system $S_1^*$ must be foreshortened by the factor of $1/\gamma_0$ so that (3) eventually attains the form

$$2\sqrt{2}\gamma_0\cos\alpha = \frac{c}{V_0} = \frac{1}{\gamma_0\beta_0} = \frac{1}{\sqrt[4]{v}},$$
$$= \frac{1}{\gamma_0} \quad (4)$$

if $\beta_0 \to 1$. This implies that an observer based at $S_1^*$ at rest will observe the Cerencov radiation from the path of the superluminally moving system $S_2'$ under a very narrow angle nearly head on if $V_0 \gg c$. As $S_2'$ propagates relative to $S_1$ with the velocity $V_0 = v_0 \gamma_0 \gg c$ the Cerenkov radiation from its flight path will arrive at $S_1^*$ from ever more distant points with decreasing frequency or increasing wavelength, respectively, and, considering the enormous velocity it is clear that it is generated practically simultaneously, as observed from the latter system. As it turns out that the order of arrival of the photons coincides with the decrease of frequency, this implies that some (arbitrary) time interval $\Delta t$ in the frequency band must also be directly coincident with the travel time of light $C\Delta t = c\gamma_0 \Delta t$. Accordingly the apparent motion of the Cerenkov image of $S_2'$ will be slowed, as is shown below.

## Application of the Relativistic Vacuum Cerenkov Effect on the Phenomenon of Gamma Ray Bursts

In the following it is proposed to apply these findings on the long standing enigma of the GRBs. Considering the introductory remarks, namely that the isotropic distribution of the GRBs on the sky leads to the rejection of the galactic halo model, we argue here that this reasoning reversely forces one to abandon the cosmic (distant galaxies) model, too, although measured red shifts in some cases seem to fully support the latter.

Suppose the cosmic model to be correct. Then, as seen from our vantage point $\approx 8.5$ kpc from the center of the Galaxy, nearly the whole galactic matter - aside from the "dark matter" - is concentrated in the central galactic bulge and in the extreme disc in the form of stars, gas and dust. Thus, a considerable amount of the gamma photons coming from point sources concealed by the galactic plane and bulge must be completely blocked by the amassed stars of the Milky Way. For photons of GRBs coming from the other side of the Milky Way especially the center of the galactic bulge must be a totally invincible hindrance. Therefore, if the cosmic model would be correct, in the direction of the galactic bulge and extreme disc, but at least the bulge, considerably less GRBs should be detectable.

The BATSE results show the contrary. Therefore, because there appears no absorption at all of GRBs by intervening material of the galactic disc, but especially the bulge (indeed, BATSE recorded at the very center of the Galaxy several GRBs, the sources of which owing to the known compact mass at this location impossibly can be located beyond the bulge) the conclusion must be drawn that the sources of GRBs are located in front of the disc and bulge, thus, ruling out any extragalactic origin. This result, the GRBs mainly to end before the galactic center and, therefore, their sources to be of galactic origin, also is an obvious constraint to any distribution of any known source candidates for GRBs within the Galaxy. None of the known astronomical objects in the Milky Way come into question as emitters of GRBs.

So, what could they be?

It is here predicted that the sources of GRBs are no known astronomical, but rather objects propagating with superluminal speed of about $\geq 10^4$ c according to (1) through the galactic space and beyond, thereby generating Cerenkov gamma and other radiation according to (2) and (4). Presumably these superluminal bodies are spaceprobes (perhaps for communication purposes) of galactic super civilisations, though, of course, the underlying physics of the propulsion and its power source remains completely unknown to say the least. In the following it is shown that the distances of the GRBs as well as the distribution of burst durations (Fig. 1) are calculable and that the main features of GRBs summarized above do exactly coincide with this hypothesis.

Consider an extraterrestrial spaceprobe hurtling in a yet unknown distance relative to Earth with a velocity of $2.637 \times 10^4$ c in accordance with the relativistic far range transformation and (1) through the galactic space. Be this so, then it must according to (2) generate Cerenkov radiation, which according to (4) is emitted along and in a very narrow cone practical in the direction of its flight path with the highest conventional photon energy to be $E_{phot} = 2$ MeV $\Rightarrow$ $E = 2$ keV. Thus, an observer in the vicinity of Earth will observe a GRB exactly then, when the spaceprobe crosses the line of sight in a very narrow angle (Fig. 2) and the generated Cerenkov radiation is bright enough to be observed. It is clear that the intensity of the Cerenkov light must strongly depend on the size and shape of the spacecraft. But at the present nothing can be said to this point and must be treated on a purely phenomenological level.

Of course, not only photons of $E = 2$ keV are generated, but a radiation cone with ones of lower energy till down to the radio waveband, too.
Suppose BATSE is detecting the GRB on all 4 channels (1, 2, 3, 4), which correspond to energy ($E_{phot}$) ranges of 20 - 50, 50 - 100, 100 - 300 and 300 - 2000 keV, respectively, with a duration of 80 s. From the top energy $E_{top} = 2$ keV according to (2) the mentioned superluminal velocity $V_0/c = \gamma_0 = 2.637 \times 10^4$ follows. In connection with the half angle of the radiation cone of the γ-photons of lowest ($E_l$) and highest ($E_h$) energy, respectively, of the respective BATSE-channel the distance of the GRB is calculable as follows:

$$d = \frac{l \times \sin\beta}{\sin\alpha}, \qquad (5)$$

where d means distance to the GRB, l = BC = length of spacecraft track - as "seen" from the vicinity of Earth in gamma light -, β the half angle of the radiation cone of the γ-photons of lowest energy and α the angle under which l = BC is to observe on the sky from the vicinity of Earth (see Fig.2).

Fig. 2 shows (exaggerated) that the "visible" gamma light track on the sky constitutes the side BC of the triangle ABC, where AB and AC are the "lines of sight" to the location of generation of the photons of lowest (B) and top (gamma) energy (C), respectively, and the distance "d" coincides with AC of this triangle. According to (4) will the Cerenkov point source not be observed to recede backward in time along BC with the steady velocity $2.637 \times 10^4$ c, but rather with ever lower velocity depending on the cosine of the Cerenkov angle of emission (Fig. 2). Obviously the relation holds

$$\frac{\cos \alpha_\gamma}{\cos \alpha_v} = \frac{V_{0_\gamma} \Delta t_\gamma}{V_{0_v} \Delta t_v},$$

which results in

$$\frac{\Delta t_\gamma}{\Delta t_v} = \frac{V_{0_v}^2}{V_{0_\gamma}^2}, \tag{6}$$

where γ denotes the respective value of higher and v the one of lower energy or frequency. Furthermore, it must be valid

$$C \Delta t_\gamma = c \gamma_0 \Delta t_\gamma = v_{app_v} \Delta t_v,$$

wherefrom in connection with (6) is derived

$$v_{app_v} = C \frac{V_{0_v}^2}{V_{0_\gamma}^2},$$
$$= c \gamma_0 \frac{\gamma_{0_v}^2}{\gamma_{0_\gamma}^2}, \tag{6a}$$

where $v_{app}$ means apparent velocity of the Cerenkov image on the sky in the light of lower frequencies (v) than gamma.

From (2) follows that the photon energy of the Cerenkov radiation tends to a maximum value if cos α ⇒ 0 implying the existence of a maximum superluminal velocity $V_{0max}$ if only cos α would attain a minimum value > 0. That is the case, because according to [8] a quantum of time exists with the numerical value $\tau_1 = 1.357628 \times 10^{-24}$ s [9]. Thus, we have

$$\cos \alpha_{min} = v_{max}^{-\frac{1}{4}} = \tau_1^{\frac{1}{4}} = 1.079 \times 10^{-6},$$

wherefrom follows $V_{0max} = 0.926 \times 10^6$ c and the highest possible Cerenkov photon energy $E_{max} = 3.43$ GeV ⇒ $E_{photmax} = 5.9 \times 10^6$ TeV. So eventually (6a) can be written as

$$v_{app_v} = C \frac{\gamma_{0_v}^2}{\gamma_{0_{max}}^2} = c \gamma_0 \frac{\gamma_{0_v}^2}{\gamma_{0_{max}}^2}. \tag{6b}$$

The Cerenkov half angle γ' follows from (4). Thus, we have γ = 180° - γ' and, therefrom, α = 180° - β - (180° - γ') = γ' - β so that (5) attains the form

$$d = \frac{l \times \sin \arccos \beta}{\sin(\arccos \gamma' - \arccos \beta)}.$$

In connection with (2), (4) and (6b) we finally obtain the GRB distance (here in light-years (lys))

$$d = \Delta t \times \gamma_{0_v} \times \frac{\dfrac{\sin \arccos \gamma_{0_v}^{-1}}{2\sqrt{2}\gamma_0}}{\dfrac{\sin\left(\arccos \gamma_{0_\gamma}^{-1} - \arccos \gamma_{0_v}^{-1}\right)}{2\sqrt{2}\gamma_0} \times 3.1536 \times 10^7} \times \left(\frac{\gamma_{0_v}}{\gamma_{0max}}\right)^2, \qquad (7)$$

where $\Delta t$ denotes duration of the GRB, $\gamma_{0_\gamma} = (E_h/h)^{1/4}$, $\gamma_{0_v} = (E_l/h)^{1/4}$ and $\gamma_0 = (E_{top}/h)^{1/4}$ - $E_h$ highest and $E_l$ lowest γ-photon energy of the respective detector channel of BATSE or other satellites (the following calculations are solely based on data from channel 2 and 3 of BATSE).

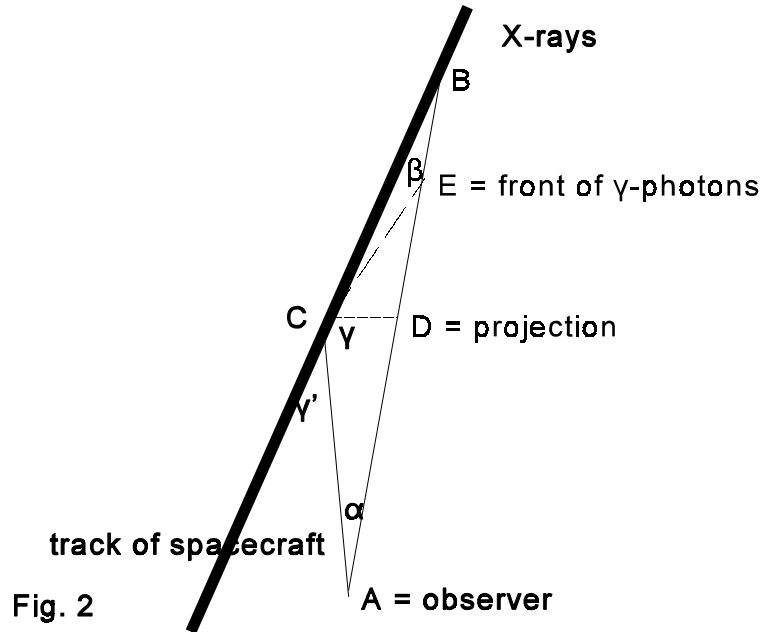

Fig. 2

Applied straightforwardly on our model spacecraft with a GRB duration of 80 s (7) delivers a distance of ≥ 4.3 lys and for some of the most energetic and longest GRBs ever detected the following values ( $E_h$ and $E_l$ taken from channel 2 (50 - 100 keV) of BATSE):

| burst duration | $E_{top}$ | E | velocity | distance |
|---|---|---|---|---|
| s | GeV | keV | $V_0/c = \gamma_0$ | lys |
| 84 | 10 | 141.42 | 76472 | ≥ 4.3 |
| 200 | 0.314 | 25.06 | 49616 | ≥ 10.7 |
| 100 | 1.2 | 48.99 | 58668 | ≥ 5.3 |
| 30 | 1.6 | 56.57 | 60816 | ≥ 1.6 |

These results reveal GRBs to be a very local phenomenon and confirm the introductory conclusion that the GRBs are confined to a space with a radius ending far before the galactic center. For the first maximum at ≈ 0.4 s in the established tri(bi)modal distribution of burst durations (No. 6 of the synopsis and Fig. 1) a distance of 7.8 light-days and for the second maximum at ≈ 1.35 s of 27.3 light-days are computed. The distance of the absolute GRB maximum at ≈ 30 s is ≥ 1.6 lys, whereas the millisecond bursts are only a couple of light-hours away.

Besides, the model GRB with a duration of 80 s coincides about with the real GRB 970228 (although in the literature I could not find any hint to the $E_{top}$ of this burst - BATSE could not observe the burst). Thus the calculated distance of ≥ 4.3 lys must approximately be valid for this burst, too. The accurate position determination of this GRB by BeppoSAX for the first time led to the discovery of a fading X-ray source and an optical counterpart of the burst. About a month after the GRB event the proper motion of the optical counterpart was detected [10]. For the angular displacement the transverse velocity-distance relation v(km/s) = 2.7 × d(pc) has been derived (Caraveo et al. 1997), where pc denotes parsec. Our theory delivers for the recession of the Cerenkov point source on the projection $\overline{CD}$ of the sky the apparent velocity (see Fig. 2 and (6b))

$$v_{app_v} = \sin\beta \times C \times \frac{\gamma_{0_v}^2}{\gamma_{0_{max}}^2}. \tag{8}$$

which gives 3,6 km/s, where $\gamma_{0v}$ = 5400, the highest posible value for the optical. The phenomenological relation also delivers a transversal velocity of 3.6 km/s for the distance of 4.3 lys of GRB 970228. Thus, theory and experiment seem to be in very good accord. But possibly the exerimental result is somewhat to high. If in (8) $\gamma_{0v}$ = 4900 - the mean $\gamma_{0v}$ of the optical - is inserted, $v_{appv}$ = 3.0 km/s results.

According to this theory the nebulosity or "fuzz" connected with the point source of the optical "afterglow" of this GRB (which has been interpreted as its host galaxy) must be streaming interstellar gas, which the superluminal spaceprobe propelled about transversal to its flight path into the interstellar space and, therewith, also normal to the line of sight from Earth to the Cerenkov track. We see the receding Cerenkov image of the spaceprobe track through this column of streaming gas. The uniform velocity of the latter explains the very small dispersion of the absorption lines (No. 12 of the synopsis). Accordingly, the fuzzy and cloudy appearance is due to the scattering of the Cerenkov light within the cloud of streaming gas. If we compare the images of the X-ray counterpart of GRB 970228 taken by BeppoSAX on February 8 and March 3 1997 [11] with the first HST observations 26 and 38 days after the burst [12] we clearly see that the fuzz is already visible in X-rays with exactly the same morphology as in the optical, although somewhat less extended. With the X-ray counterpart the "X-ray fuzz" vanished, too, and it is predicted that the optical nebulosity will share the same fate, which implies that it disappeared already (as well as the optical point source) since the last HST observation in september 1997. This conclusion is further backed by the comparison of ground-based measurements taken on March 6 and April 5 and 6, which clearly indicate that within a month the fuzz faded below detection level of the telescope [13]. From the foregoing it is clear that the observed red shifts of the afterglow of some GRBs are not of cosmological, but rather special relativistic origin, where the measured red shift is but only the imprint of the transversal Doppler-shift $z = \gamma_0 - 1$. This implies all red shift measurements other than from the putative afterglows to be erronous.

From the theory it is clear that repeaters do not exist (although recurrence cannot be excluded if those spaceprobes fly the same routes in some time interval) and why all attempts to associate the GRBs with known astronomical objects will be futile (see below). This implies that the identification of "host galaxies" as e. g. in the case of GRB 971214 is erronous and must be due to the chance superposition of the GRB respectively its afterglow and the putative host.

Furthermore, follows that the random distribution of GRBs on the sky results from the accidental crossing of the line of sight by spacecraft or -probes criss-crossing the galactic space in all possible directions relative to Earth with superluminal velocities exceeding $\approx 10^4$ c. From this directly results that far more though invisible (from Earth) spacecraft tracks in Cerenkov gamma light must be generated than the stated 0,8/day. As already mentioned, with growing time Cerenkov radiation of ever lower frequency will be observed from the spaceprobe track. Thus, an observer really sees the evolution of the track in Cerenkov light of ever lower frequency backward in time (see below). This is valid for all GRBs. But in the case of very short bursts the duration of the following Cerenkov radiation of lower energy is also proportionally shortened and therefore hard to detect (No. 11 of the synopsis). According to (6) can the duration time of the Cerenkov afterglow at any frequency be approximated by the expression (see Fig. 2):

$$\Delta t_\nu = \Delta t_\gamma \frac{V_{0_\gamma}^2}{V_{0_\nu}^2} \qquad (9)$$

where $\Delta t_\nu$ means duration of the respective waveband (X-ray, optical, radio) and $\Delta t_\gamma$ duration of GRB. Applied on GRB 970228 and GRB 970508 (9) delivers about the correct results of some days for the Cerencov X-ray glow and a couple of months up to

≥ one year for the visual Cerenkov light.

## GRB Properties as a Consequence of the Track Geometry

Due to the extreme beaming of the Cerenkov radiation without any deflection whatsoever from its generation by nonhuman made superluminal spaceprobes till its detection by human made satellites, the gamma light does not expand isotropically but cylindrically around the "visible" gamma track. Because the "hight" of the "cylinder" l = BC (Fig. 2) does not alter on the way of the photons to the detector, only the "cylinder barrel" expands with the superluminal velocity C of light. This implies that the intensity does not decrease as $1/d^2$ - as in the case of normal astronomical objects - but rather as $1/d$ instead. Therefore, according to this theory, the intensity of GRBs is exactly dependent on their distance, which again is proportional to the burst duration. If the mean of the durations (distances) of the longest lasting GRBs (table) is divided by the duration (distance) of the shortest bursts of about 0.01 s, 10350 results as compared with the empirical value ≈ 10000 (No. 10 of the synopsis).

These findings suggest that the absolute maximum of GRB durations at ≈ 30 s ⇒ d ≈ 1.6 lys and their rapid decline is not real but faked by the pesumable smallness of the sources and the limited sensitivity of current detectors, analogous to the limited number of stars visible to the naked eye. This implies that the maxima at ≈ 0.4 s ⇒ d ≈ 7.8 light-days and ≈ 1.35 s ⇒ d ≈ 27.3 light-days owe their existence to a real increase of superluminal space travel activities in these distances (this result will be confirmed below). Thus, it can be said that the limitation for the detection of GRBs due to the geometry of the spaceprobe tracks is further and decisive reduced by the fact that beyond burst durations of Δt ≈ 30 s ⇒ d ≈ 1.6 lys the GRBs drop rapidly below the range of visibleness.

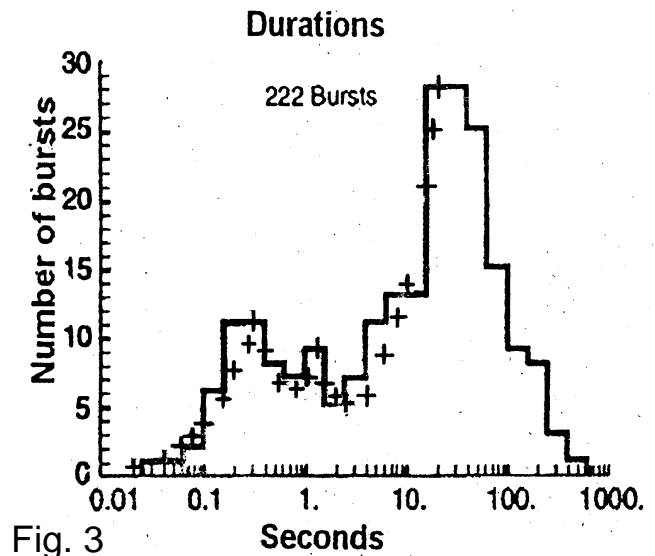

Fig. 3

The duration or distance of the GRBs also is the key of understanding the noteworthy trimodal distribution of burst durations (No. 6. of the synopsis and Fig. 1). From Fig. 1 and (6) it is clear that at a given $E_{top}$ the burst duration is a function of the distance (and vice versa), whereas the length of the spaceprobe track l = BC can be thought of as the projection $\overline{CD}$ onto a circle, the radius of which is the distance d (see Fig. 2). Consequently, under the provision that GRBs occur randomly in space, the probability of bursts to be observed from the vicinity of Earth must be proportional to the circumference of the said circle or its radius = distance d. Therefore, the number of GRBs must be directly dependent on their distance d being equivalent to the duration time Δt. Applied on Fig. 1 this means that the number of GRBs must steadily decrease toward our vantage point in the system of the sun from the absolute maximum of 29 bursts in d ≈ 1.6 lys or burst duration ≈ 30 s, proportional to shorter burst times or distances. To the resulting curve the difference to the local maxima at Δt ≈ 1.35 s ⇒ d ≈ 27.3 light-days (11 - 29 × 1.35/30) and Δt ≈ 0.4 s ⇒ d ≈ 7.8 light-days (9 - 29 × 0.40/30) have to be added. The steady increase of GRBs toward both local maxima follows the same law, but other than the apparent decrease of GRBs beyond the absolute maximum owing to the drop beneath the range of visibleness, those local maxima must also decrease proportional to the burst duration or increasing distance. The crosses in Fig. 3 mark this theoretical derivation.

The result that the small maxima at burst duration time Δt ≈ 0.4 s ⇒ d ≈ 7.8 light-days and Δt ≈ 1.35 s ⇒ d ≈ 27.3 light-days are due to a real increase of bursts or superluminal space travel activities in the vicinity of the sun is strongly supported by the finding short GRBs to be distributed anisotropically (No. 6 of the synopsis). Especially has been found (Balázs et al. 1998) that GRBs with duration times < 2 s and <1 s, respectively, depart significantly from isotropy, whereas long bursts > 2 s do not.

As already mentioned, it is from the track geometry (Fig. 2) clear that the observer in A at first will receive γ-photons of top and subsequently of ever lower energy, which effect also causes the observed delay of low energy γ-photons (No. 9 of the synopsis). Thus, each waveband must have its own characteristic peak, which leads in the γ-region to the observed FRED phenomenon, and it is clear that this is valid for the following X-ray, optical and radio Cerenkov emission as well, with a rise to a plateau value and a subsequent decay. But this smooth evolution of the Cerenkov track can only be valid if the track geometry remains unaltered. If the spaceprobe changes its course anywhere during the "visible" track, then this also will lead to a drastic change in the evolution of the subsequent Cerenkov radiation. Suppose a spaceprobe moves on a heading, which allows the Cerenkov radiation to be observed from Earth. At a point of its track, wherefrom visual light or radio emission would reach the observer, it turns slightly to a new course leading away from Earth - comparable to an ascending aircraft. Due to this turn in the track the beam of the more energetic (than visual light or radio) Cerencov radiation arrives earlier at the observer as in the normal case with no alteration of the flight path (see Fig. 2). Thus, the observer would register the incoming γ-rays, X-rays and visual light nearly simultaneously. Possibly due to the turn the point of the Cerenkov track with γ-rays of medium energy or X-rays will be nearest to the observer with the consequence that this radiation arrives first (pre-cursor), whereas the most energetic photons appear delayed. Obviously was GRB 990123 such an event. It was observed in optical during the course of the GRB itself, yet the most energetic (MeV) emission did not rise significantly until 18 s after BATSE's lower energy burst onset [14].

The rapid fluctuations in the ms region (No. 7 of the synopsis) are presumably due to the basically stochastic nature of the Cerenkov gamma radiation. From (2) directly follows that the probability for the generation of photons of top energy $E_{top}$ (as compared with photons of lowest energy $E_l$) as a function of the velocity $V_0$ varies as

$$\frac{E_{top}}{E_l} = \left(\frac{V_0}{V_{0_l}}\right)^4 .$$

Thus, with increasing energy $E_{top}$ the time difference between two gamma photon generating events is growing and it comes to the observed fluctuations. This basically also must be true at the onset of the Cerenkov radiation of any frequency.
From the theory it is clear that the more distant a spacecraft track, the longer and smoother the "visible" GRB with ever lower intensity

### Further Implications

It seems highly improbable that all spaceships of the unknown galactic super civilisations propagate so fast through the near vacuum of the Milky Way and its vicinity that always Cerenkov gamma radiation is generated along their flight paths. More likely also spacecraft cruise the galactic space with velocities sufficing to generate photons with the top energy $E_{top}$ in the optical, ultraviolett or X-ray region. Probably are most if not all transient optical phenomena found on archival astronomical photographic plates such Cerenkov events in visual light.

In the extreme ultra-violet (EUV) waveband transients have been observed, too. For instance revealed ROSAT WFC observations a transient EUV source with a flux amplitude variability by a factor of ≈ 7000 over a maximum period of presumably ≈ one year, which has not been found in earlier EUV observations nor in follow-up γ-ray, X-ray, optical and radio observations. If (7) is straightforwardly applied on this EUV transient (date of observation: 25.06. - 07.07.1997; energy: 62 - 110 eV) this results (Δt = 13 days) in a minimum distance of 2538 lys.
Short-duration X-ray transients, which are not related to known astronomical objects, have also been observed long since. Typically they range from seconds to less than few hours [15]. BeppoSAX observed at least 9 fast X-ray bursts during its operation life time, which are not counterparts of GRBs. Nevertheless they show spectral characteristics typical of the GRB class, but do not present relevant gamma emission [16]. If e. g. (7) is applied on the WATCH fast X-ray transient GRS 2037-404 with a duration of 110 min in the 8 - 15 keV band (Tirado et. al 1999) a distance of 155 lys is calculated.

From the foregoing it is clear, especially because $E_{max}$ = 3.43 GeV ⇒ $E_{photmax}$ = 5.9 × $10^6$ TeV far exceeds the top energy of the presently observable most energetic GRBs, that we also have to expect burst phenomena in the TeV range - although presumably far more seldom than at lower energies
.
It is predicted that all these transient phenomena in the different wavebands are of common origin: Vacuum Cerenkov radiation due to the extreme superluminal propagation of interstellar spaceships or -probes of various extraterrestrial civilisations in our Galaxy.

Dedicated to my friend Tom Witte.

---